\documentclass[journal]{IEEEtran}

\usepackage{graphicx}
\usepackage{ulem}
\usepackage{subfigure} 
\usepackage{epsfig}

\begin{document}
%
% paper title
\title{Sensitivity of Antenna Arrays for Long-Wavelength Radio Astronomy}
%
% author names and IEEE memberships
% note positions of commas and nonbreaking spaces ( ~ ) LaTeX will not break
% a structure at a ~ so this keeps an author's name from being broken across
% two lines.
% use \thanks{} to gain access to the first footnote area
% a separate \thanks must be used for each paragraph as LaTeX2e's \thanks
% was not built to handle multiple paragraphs
\author{S.W.~Ellingson,~\IEEEmembership{Senior~Member,~IEEE} %<-this % stops a space
%        T.E.~Clarke,
%        A.~Cohen,
%        J.~Craig,~\IEEEmembership{Member,~IEEE}, %<-this % stops a space
%        N.E.~Kassim,
%        Y.~Pihlstr\"{o}m,
%        L.~J Rickard,
%        and G.B.~Taylor
%}% <-this % stops a space
\thanks{
S.W. Ellingson is with the Bradley Dept.\ of Electrical \& Computer Engineering, Virginia Polytechnic Institute \& State University, Blacksburg, VA 24061 USA (e-mail: ellingson@vt.edu).  
%Y.\ Pihlstr\"{o}m and G.B. Taylor are with the Dept. of Physics \& Astronomy, University of New Mexico, Albuquerque, NM 87131 USA (e-mail: gbtaylor@unm.edu).
%N.E. Kassim, T.E. Clarke, and A. Cohen are with the U.S. Naval Research Laboratory, Washington, DC 20375 (e-mail: namir.kassim@nrl.navy.mil).
%J. Craig and L. J Rickard are with the LWA Project Office, University of New Mexico, Albuquerque, NM 87131 USA (e-mail: lrickard@unm.edu).
}
}
% note the % following the last \IEEEmembership and also the first \thanks - 
% these prevent an unwanted space from occurring between the last author name
% and the end of the author line. i.e., if you had this:
% 
% \author{....lastname \thanks{...} \thanks{...} }
%                     ^------------^------------^----Do not want these spaces!
%
% a space would be appended to the last name and could cause every name on that
% line to be shifted left slightly. This is one of those "LaTeX things". For
% instance, "A\textbf{} \textbf{}B" will typeset as "A B" not "AB". If you want
% "AB" then you have to do: "A\textbf{}\textbf{}B"
% \thanks is no different in this regard, so shield the last } of each \thanks
% that ends a line with a % and do not let a space in before the next \thanks.
% Spaces after \IEEEmembership other than the last one are OK (and needed) as
% you are supposed to have spaces between the names. For what it is worth,
% this is a minor point as most people would not even notice if the said evil
% space somehow managed to creep in.
%
% The paper headers
\markboth{IEEE TRANSACTIONS ON ANTENNAS AND PROPAGATION
%,~Vol.~x, No.~x,~Month~YYYY
}{Shell \MakeLowercase{\textit{et al.}}: Sensitivity of Antenna Arrays for Long-Wavelength Radio Astronomy}
% The only time the second header will appear is for the odd numbered pages
% after the title page when using the twoside option.
% 
% *** Note that you probably will NOT want to include the author's name in **% *** the headers of peer review papers.                                   ***

% If you want to put a publisher's ID mark on the page
% (can leave text blank if you just want to see how the
% text height on the first page will be reduced by IEEE)
%\pubid{0000--0000/00\$00.00~\copyright~2009 IEEE}

% use only for invited papers
%\specialpapernotice{(Invited Paper)}

% make the title area
\maketitle

\begin{abstract}
A number of new and planned radio telescopes will consist of large arrays of low-gain antennas operating at frequencies below 300~MHz.  In this frequency regime, Galactic noise can be a significant or dominant contribution to the total noise.  This, combined with mutual coupling between antennas, makes it difficult to predict the sensitivity of these instruments.  This paper describes a system model and procedure for estimating the system equivalent flux density (SEFD) -- a useful and meaningful metric of the sensitivity of a radio telescope -- that accounts for these issues.  The method is applied to LWA-1, the first ``station'' of the Long Wavelength Array (LWA) interferometer.  LWA-1 consists of 512 bowtie-type antennas within a 110 $\times$ 100~m elliptical footprint, and is designed to operate between 10~MHz and 88~MHz using receivers having noise temperature of about 250~K.  It is shown that the correlation of Galactic noise between antennas significantly desensitizes the array for beam pointings which are not close to the zenith.  It is also shown that considerable improvement is possible using beamforming coefficients which are designed to optimize signal-to-noise ratio under these conditions.  Mutual coupling is found to play a significant role, but does not have a consistently positive or negative influence.  In particular, we demonstrate that pattern multiplication (assuming the behavior of single antennas embedded in the array is the same as those same antennas by themselves) does not generate reliable estimates of SEFD.      
%The Long Wavelength Array (LWA) will be a new multi-purpose radio telescope operating in the frequency range 10--88 MHz. Upon completion, LWA will consist of 53 phased array ``stations'' distributed over a region about 400~km in diameter in the state of New Mexico. Each station will consist of 256 pairs of dipole-type antennas whose signals are formed into beams, with outputs transported to a central location for high-resolution aperture synthesis imaging.  The resulting image sensitivity is estimated to be a few mJy ($5\sigma$, 8~MHz, 2 polarizations, 1 hr, zenith) in 20--80~MHz; with resolution and field of view of ($8''$,$8^{\circ}$) and ($2''$,$2^{\circ}$) at 20~MHz and 80~MHz, respectively.  Notable engineering features of the instrument, demonstrated in this paper, include Galactic-noise limited active antennas and direct sampling digitization of the entire tuning range. This paper also summarizes the LWA science goals, specifications, and analysis leading to top-level design decisions.
\end{abstract}

% Note that keywords are not normally used for peerreview papers.

% For peer review papers, you can put extra information on the cover
% page as needed:
% \begin{center} \bfseries EDICS Category: 3-BBND \end{center}
%
% For peerreview papers, inserts a page break and creates the second title.
% Will be ignored for other modes.
\IEEEpeerreviewmaketitle

%========================================================
\section{\label{sIntro}Introduction}
%========================================================

A number of new and planned radio telescopes will consist of large arrays of closely-spaced low-gain antennas operating at frequencies below 300~MHz.  These include LWA \cite{PIEEE_LWA}, LOFAR \cite{PIEEE_LOFAR},  MWA \cite{PIEEE_MWA}, and SKA \cite{PIEEE_SKA}.  In this frequency regime, it is possible to design receivers with noise temperatures that are much less than the antenna temperatures associated with the ubiquitous Galactic synchrotron radiation, such that the resulting total system noise temperature is dominated by Galactic noise \cite{E05}.  This is quite different from the condition most often considered, in which it is usually assumed that internal noise associated with the receivers dominates and does not scatter into the array, so that the noise associated with different antennas is uncorrelated; see e.g. \cite{VB88,Lee93,Kraft00}.  Recent work on arrays for long-wavelength radio astronomy accounts for the dominance of external noise, but neglects the effects of correlation of this noise between antennas \cite{VBA05,PIEEE_LWA}.  However, it is known that this correlation is likely to have a significant effect for these instruments \cite{LWA142}, and this issue is further explored in this paper.  In \cite{CPD04,C05}, and \cite{IMW08}, correlation of noise between antennas is considered, but the source of noise is internal (amplifiers or ohmic losses in antennas) and the correlation arises due to propagation internal to the array.  The problem of correlation of external noise has been intermittently considered in communications (e.g. \cite{Gans06}) and direction finding (e.g. \cite{AST07}) applications, but does not seem to have been previously considered for long-wavelength radio astronomy beamforming applications.      

Because antenna spacing in the systems of interest is typically less than a few wavelengths, mutual coupling also plays a significant role.  Since these arrays are electromagnetically large, interact with the electromagnetically-complex ground, and may have aperiodic spacings, it is difficult to determine the characteristics of antennas, either individually or collectively as part of a beamforming system.  
%Thus, the {\it array manifold} that is, the response of the array to a plane wave incident from each possible direction of interest -- is  
In particular, it is usually difficult to know if pattern multiplication -- that is, assuming that the behavior of single antennas embedded in the array is the same as those same antennas by themselves -- yields reasonable results.  Past studies have shown that mutual coupling in aperiodic arrays of low-gain elements results in fluctuation of beam gain and sidelobe levels as a function of scan angle when element spacing is less than a few wavelengths 
%-- spacings which are typical for the systems of interest 
\cite{LS67,AL72}.  This suggests that pattern multiplication may not be a useful assumption.  
%The mutual coupling issue also further complicates calculation of the effects of correlation of Galactic noise between antennas.             
However, useful and generalizable findings which are applicable to the systems of interest are not commonly available. 

This paper describes a procedure for estimating the sensitivity of radio telescope arrays which is appropriate under these conditions.  The procedure is based on a system model, described in Section~\ref{sTheory}, which relates the electromagnetic response of the array (the {\it array manifold}), a model for the external noise temperature, and a model for the receiver noise temperatures to the system equivalent flux density (SEFD) achieved by a beam formed using specified beamforming coefficients.  
%The array manifold is defined as the voltage generated across the antenna terminals of all antennas in the array, terminated into the input impedance offered by the receivers, in response to a plane wave incident from a specified direction, for all possible directions of incidence.
SEFD is defined as the power flux spectral density (i.e., W~m$^{-2}$~Hz$^{-1}$) which yields signal-to-noise ratio (SNR) equal to unity at the beamformer output.  SEFD is a useful metric as it includes the combined effect of antennas and all noise sources into a single ``bottom line'' number that is directly related to the sensitivity of astronomical observations.       

The primary difficulty in determining SEFD using the above procedure is obtaining the array manifold.  One approach is demonstrated by example in Section~\ref{sLWA} of this paper.  We analyze LWA-1, the first ``station'' of the Long Wavelength Array (LWA) interferometer \cite{PIEEE_LWA}.  LWA-1 consists of 512 bowtie-type antenna elements arranged into 256 dual-polarized ``stands'' within a 110 $\times$ 100~m elliptical footprint.  LWA-1 is designed to operate between 10~MHz and 88~MHz using receivers having noise temperature of about 250~K.  We obtain the array manifold for LWA-1 at 20~MHz, 38~MHz, and 74~MHz using a method of moments (MoM) wire-grid model.  Because the model is too large to analyze all at once (a common problem with this class of arrays), a procedure described in Section~\ref{ssArray} is employed in which the manifold is calculated for one stand (i.e., one pair of collocated antenna elements) at a time.  Unlike pattern multiplication in which the presence of the remaining stands would be ignored, this procedure obtains the response of each antenna in the presence of nearby antennas and structures.                

%An additional useful aspect of the approach described in Sections~\ref{sTheory} and \ref{sLWA} is that is provides a method for determining
Also considered in this paper is the selection and performance of beamforming coefficients which optimize SEFD.  Because mutual coupling and external noise correlation are significant, it is to be expected that ``simple'' beamforming coefficients based solely on antenna positions (i.e., phases associated with geometry only) will not be optimal.  In Section~\ref{sAP}, the SEFD performance of LWA-1 is evaluated.  It is shown that the optimal coefficients significantly improve sensitivity relative to simple coefficients.  Finally, in Section~\ref{sConc}, the extension of these results to predict the imaging performance of an interferometer comprised of multiple beamforming arrays is considered.

%========================================================
\section{\label{sTheory}Theory}
%========================================================

Let $E_{\theta}(t)$ and 
    $E_{\phi  }(t)$ be the 
    $\theta$- and $\phi$-polarized components of the electric field of the signal of interest, 
    having units V~m$^{-1}$~Hz$^{-1/2}$. 
In this coordinate system, $\theta$ is measured from the $+z$ axis, which points toward the zenith; the ground lies in the $z=0$ plane, and $\phi$ is measured from the $+x$ axis.
The signal of interest is incident from $\{ \theta_0,\phi_0 \}$, which is henceforth indicated as $\psi_0$.  
The resulting voltage across the terminals of the $n^{\mbox{th}}$ antenna element, having units of V~Hz$^{-1/2}$, is
\begin{equation}
x_n(t) = a^{\theta}_n(\psi_0) E_{\theta}(t)
               +a^{\phi  }_n(\psi_0) E_{\phi  }(t)
               +z_n(t) + u_n(t)
\label{exn}
\end{equation}
where: $a^{\theta}(\psi_0)$ and $a^{\phi}(\psi_0)$ are the effective lengths, having units of meters, associated with the $\theta$ and $\phi$ polarizations, respectively, for the $n^{\mbox{th}}$ antenna element for signals incident from $\psi_0$; $z_n(t)$ is the contribution from noise {\it external} to the system; and $u_n(t)$ is the contribution from noise {\it internal} to the system.  Note $u_n(t)$ can also include internal noise unintentionally radiated by some other antenna and received by antenna $n$.  In all cases, we assume these quantities are those which apply when antennas are terminated into whatever electronics are actually employed in the system, as opposed to being ``open circuit'' or ``short circuit'' quantities.   Without loss of generality we can interpret these to be time-harmonic (i.e., monochromatic complex-valued ``baseband'') quantities.  

Beamforming can be described as the operation:
\begin{equation}
y(t) = \sum_{n=1}^{N}{ b_n x_n(t) }
\label{ey}
\end{equation}
where $N$ is the number of antennas, and the unitless $b_n$'s specify the beam.  Assuming root-mean-square voltages, the power at the output of the beamformer is 
\begin{equation}
P_y = \left< y(t) y^*(t) \right> R_{o}^{-1}
\label{ePy}  
\end{equation}  
where $<\cdot>$ denotes time-domain averaging and ``$^*$'' denotes conjugation, and $R_o$ is the impedance looking into the system as seen from the terminals across which $y(t)$ is measured, assumed to be purely resistive.  

We now wish to evaluate Equation~\ref{ePy} by substitution of Equation~\ref{ey}.  In the process of expanding Equation~\ref{ePy}, let us assume that the signal of interest, $z_n(t)$, and $u_n(t)$ are mutually uncorrelated for any given $n$.  Specifically, we assume that for any $n$ and $m$:
\begin{equation}
\left< E_{\theta}(t) z_n^*(t) \right> = \left< E_{\phi}(t) z_n^*(t) \right> = 0
\end{equation}
\begin{equation}
\left< E_{\theta}(t) u_n^*(t) \right> = \left< E_{\phi}(t) u_n^*(t) \right> = 0
\end{equation}
\begin{equation}
\left< z_n(t) u_m^*(t) \right> = 0 ~\mbox{.}
\end{equation}
Note that the possibility that like terms are correlated between antennas is not precluded by the above assumptions; for example, $\left< z_n(t) z_m^*(t) \right>$ can be $\neq 0$ for $n \neq m$.  Furthermore, we have not yet made any assumption about the correlation between $E_{\theta}(t)$ and $E_{\phi}(t)$. Under these assumptions, Equation~\ref{ePy} can be written as follows:
\begin{eqnarray}
P_y R_o  = & & {\bf b}^H {\bf A}_{\theta\theta} {\bf b}~P_{\theta\theta} + {\bf b}^H {\bf A}_{\phi\phi  } {\bf b}~P_{\phi  \phi} \nonumber \\
         &+& {\bf b}^H {\bf A}_{\theta\phi  } {\bf b}~P_{\theta\phi  } + {\bf b}^H {\bf A}_{\phi\theta} {\bf b}~P_{\phi\theta} \nonumber \\
         &+& {\bf b}^H {\bf P}_z              {\bf b} ~~~~~~~  + {\bf b}^H {\bf P}_u            {\bf b}
~\mbox{,}
\label{ePy1} 
\end{eqnarray}
where ``$^H$'' denotes the conjugate transpose operator;
\begin{equation}
{\bf b} = \left[ ~ b_1 ~ b_2 ~ \cdots ~ b_N ~ \right]^T
~\mbox{,}
\end{equation}
where ``$^T$'' denotes the transpose operator; and
\begin{equation}
{\bf A}_{\theta\theta} = {\bf a}_{\theta}^*(\psi_0) ~ {\bf a}_{\theta}^T(\psi_0)  ~~~ P_{\theta\theta} = \left< \left| E_{\theta}(t) \right|^2 \right>~~
\end{equation}
\begin{equation}
{\bf A}_{\phi  \phi} = {\bf a}_{\phi}^*(\psi_0) ~ {\bf a}_{\phi  }^T(\psi_0)  ~~~ P_{\phi  \phi  } = \left< \left| E_{\phi}(t) \right|^2 \right>~~
\end{equation}
\begin{equation}
{\bf A}_{\theta\phi} = {\bf a}_{\phi}^*(\psi_0) ~ {\bf a}_{\theta}^T(\psi_0)  ~~~ P_{\theta\phi  } = \left< E_{\theta}(t) E_{\phi}^*(t) \right>
\end{equation}
\begin{equation}
{\bf A}_{\phi\theta} = {\bf a}_{\theta}^*(\psi_0) ~ {\bf a}_{\phi}^T(\psi_0) ~~~ P_{\phi\theta  } = \left< E_{\phi}(t) E_{\theta}^*(t) \right>
\end{equation}
\begin{equation}
{\bf a}_{\theta}(\psi_0) = \left[ ~ a_1^{\theta}(\psi_0) ~ a_2^{\theta}(\psi_0) ~ \cdots ~ a_N^{\theta}(\psi_0) ~ \right]^T
\end{equation}
\begin{equation}
{\bf a}_{\phi  }(\psi_0) = \left[ ~ a_1^{\phi  }(\psi_0) ~ a_2^{\phi  }(\psi_0) ~ \cdots ~ a_N^{\phi  }(\psi_0) ~ \right]^T
~\mbox{;}
\end{equation}
and
${\bf P}_z$ is a matrix whose $(n,m)^{\mbox{th}}$ element is $\left< z_n^*(t) z_m(t) \right>$, and 
${\bf P}_u$ is a matrix whose $(n,m)^{\mbox{th}}$ element is $\left< u_n^*(t) u_m(t) \right>$.\footnote{Following the convention that first index indicates row and the second index indicates column.}

We now consider the external noise correlation matrix ${\bf P}_z$.  We wish to obtain a simple expression for ${\bf P}_z^{\left[n,m\right]}$, the $\left(n,m\right)^{\mbox{th}}$ element of ${\bf P}_z$, in terms of physical quantities more relevant to radio astronomy.  First, let us define $\Delta S(\psi)$ as the flux density, having units of W~m$^{-2}$~Hz$^{-1}$, associated with the electric field $\Delta E(\psi,t)$ incident from a region of solid angle $\Delta\Omega$ around $\psi$.  Assuming $\Delta E(\psi,t)$ is given in terms of root-mean-square voltage, the relationship is 
\begin{equation}
\Delta S(\psi) = \left< \left| \Delta E(\psi,t) \right|^2 \right> / \eta
\end{equation}  
where $\eta$ is the impedance of free space.  
%Neglecting anthropogenic sources, 
Since the Galactic synchrotron background noise is essentially unpolarized, we assume the powers in the $\theta$- and $\phi$-polarized components of $\Delta E(\psi,t)$ are equal; specifically,
\begin{equation}
  \left< \left| \Delta E_{\theta}(\psi,t) \right|^2 \right>  
= \left< \left| \Delta E_{\phi  }(\psi,t) \right|^2 \right>  =  \frac{\eta}{2} \Delta S(\psi)
\end{equation} 
Note also that $\Delta S(\psi)$ can be obtained independently from the Rayleigh-Jeans Law:
\begin{equation}
\Delta S(\psi) = \frac{2k}{\lambda^2} T_e(\psi) \Delta\Omega
\end{equation}  
where $k$ is Boltzmann's constant ($1.38 \times 10^{-23}$~J/K), $T_e(\psi)$ is the apparent external noise brightness temperature (attributable to either Galactic noise or thermal radiation from the ground) in the direction $\psi$, and $\lambda$ is wavelength.  We can model $\Delta E_{\theta}(\psi,t)$ and $\Delta E_{\phi}(\psi,t)$ as follows:
\begin{equation}
\Delta E_{\theta}(\psi,t) = g_{\theta}(\psi,t) \sqrt{ \frac{k\eta}{\lambda^2} T_{e}(\psi) \Delta\Omega }  
\end{equation}  
\begin{equation}
\Delta E_{\phi  }(\psi,t) = g_{\phi  }(\psi,t) \sqrt{ \frac{k\eta}{\lambda^2} T_{e}(\psi) \Delta\Omega }  
\end{equation} 
where $g_{\theta}(\psi,t)$ and $g_{\phi}(\psi,t)$ are Gaussian-distributed random variables with zero mean and unit variance.  Note that we expect not only that $g_{\theta}(\psi,t)$ and $g_{\phi}(\psi,t)$ will be independent random variables, but also that $g_{\theta}(\psi_1,t)$ and $g_{\theta}(\psi_2,t)$ will be uncorrelated for $\psi_1 \neq \psi_2$, and similarly for $g_{\phi}(\psi,t)$.  We obtain $z_n(t)$ by summing up the contributions received over a sphere:\footnote{Writing this as a discrete sum as opposed to an integral yields a general result while avoiding the complication of fractional calculus.}
\begin{equation}
z_n(t) = \sum_{\psi}{ \left[ a^{\theta}_n(\psi) \Delta E_{\theta}(\psi,t) 
                                   + a^{\phi  }_n(\psi) \Delta E_{\phi  }(\psi,t) \right] }
\end{equation}
Applying the definition of ${\bf P}_z$ and exploiting the statistical properties of $g_{\theta}(\psi,t)$ and $g_{\phi}(\psi,t)$, we find:
\begin{equation}
{\bf P}_z^{\left[n,m\right]} = \\
\frac{k\eta}{\lambda^2} \sum_{\psi}{ 
\left[ a^{\theta *}_n(\psi) a^{\theta}_m(\psi)
     + a^{\phi   *}_n(\psi) a^{\phi  }_m(\psi) \right] 
T_{e}(\psi) \Delta\Omega }
\end{equation}
which can now be written in integral form:
\begin{equation}
{\bf P}_z^{\left[n,m\right]} = \\
\frac{k\eta}{\lambda^2} \int_{}{  
\left[ a^{\theta *}_n(\psi) a^{\theta}_m(\psi)
     + a^{\phi   *}_n(\psi) a^{\phi  }_m(\psi) \right] 
T_{e}(\psi) d\Omega}
\label{ePzInt}
\end{equation}

Returning to Equation~\ref{ePy1}, note that the signal to noise ratio (SNR) at the output of the beamformer can be written as:
\begin{equation}
\mbox{SNR} = 
                                \frac{  {\bf b}^H {\bf R}_s {\bf b} }
                                     {  {\bf b}^H {\bf R}_n {\bf b} }~\mbox{,}
\label{eSNR}
\end{equation}
where
\begin{equation}
{\bf R}_s = {\bf A}_{\theta\theta} P_{\theta\theta} 
           + {\bf A}_{\phi  \phi  } P_{\phi  \phi  } 
           + {\bf A}_{\theta\phi  } P_{\theta\phi  } 
           + {\bf A}_{\phi  \theta} P_{\phi  \theta}~\mbox{, and}
\end{equation}
\begin{equation}
{\bf R}_n = {\bf P}_z + {\bf P}_u~\mbox{.}
\end{equation}

In general, the maximum possible SNR is equal to the maximum eigenvalue of ${\bf R}_n^{-1}{\bf R}_s$, and is achieved by selecting ${\bf b}$ to be the corresponding eigenvector \cite{MM80} (see also \cite{VB88}).  Alternative approaches to beamforming include (1) selecting ${\bf b} = {\bf a}_{\theta}^*(\psi_0)$ or ${\bf a}_{\phi}^*(\psi_0)$, which accounts for mutual coupling but neglects spatial noise correlation; or (2) selecting the beamforming coefficients to compensate only for the geometrical delays, which neglects both effects.  Approach (1) is optimal when 
%the incident electric field is linearly-polarized and 
the noise associated with each antenna is uncorrelated, such that 
${\bf R}_n$ has the form $\sigma_n^2{\bf I}$; i.e., some constant times the identity matrix.  Then, we obtain the well-known result that the SNR improves linearly with $N$.  However, as will be demonstrated in later in this paper, this special case is not necessarily relevant to the problem of interest.  This is primarily due to the impact of the external noise correlation, as represented by ${\bf P}_z$, for which off-diagonal terms can be significant.     

For a signal of interest which is unpolarized -- a useful and common assumption for the purpose of characterizing the sensitivity of a radio telescope -- we have $P_{\theta\phi} = P_{\phi\theta} = 0$ and $P_{\theta\theta} = P_{\phi\phi} = \eta S(\psi)/2$, where $S(\psi)$ is the flux (distinct from the external noise considered above) associated with the signal of interest.  In this case, we have
\begin{equation}
{\bf R}_s =  S(\psi) \frac{\eta}{2} \left(  {\bf A}_{\theta\theta} + {\bf A}_{\phi\phi} \right)~\mbox{.}
\label{eRs}
\end{equation}
Continuing with this assumption, it is convenient to express the sensitivity of a radio telescope in terms of {\it system equivalent flux density} (SEFD), defined as the value of $S(\psi)$ in Equation~\ref{eRs} required to double the total power observed at the beamformer output; i.e., SNR$=$1.  Thus,
\begin{equation}
\mbox{SEFD} = \frac{2}{\eta}
              \frac{  {\bf b}^H \left( {\bf P}_z + {\bf P}_u \right) {\bf b} }
                   {  {\bf b}^H \left(  {\bf A}_{\theta\theta} + {\bf A}_{\phi\phi} \right) {\bf b} }~\mbox{.}
\label{eSEFD}
\end{equation}
This expression is useful as it describes the sensitivity of a radio telescope array in terms of the array manifold, the contributions of internal and external noise, and the beamforming coefficients.  The principal difficulty in using this expression is calculating the array manifold.  This is considered next. 

%========================================================
\section{\label{sLWA} LWA-1 Design and Array Manifold}
%========================================================  

\begin{figure}
\begin{center}
%\vspace{3in}
%\psfig{file=figs/Station.eps,width=3.1in}
\psfig{file=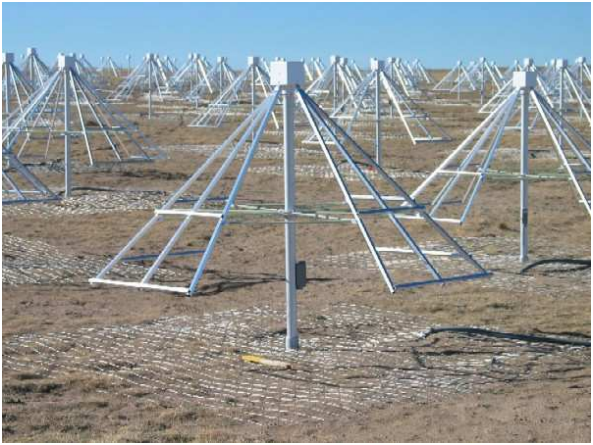,width=3.1in}
\end{center}
\caption{
\label{fStation}
LWA-1 under construction (picture taken November 2009), showing a few of the completed stands. 
}
\end{figure}

The original motivation behind the work presented in this paper was to characterize the performance of LWA-1. LWA-1 consists of $N=512$ antennas arranged into 256 ``stands'', with each stand consisting of two orthogonally-aligned bowtie-type dipoles over a wire mesh ground screen, as shown in Figure~\ref{fStation}.  In Sections~\ref{ssSingleStand} and \ref{ssArray} we review the relevant details of the design of the LWA-1 array, our method for computing the array manifold and the internal and external noise covariance matrices, and present some results.  

%--------------------------------------------------------
\subsection{\label{ssSingleStand}LWA Stand Design \& Electromagnetic Model}

As the dipoles and ground screen comprising each stand consist of interconnected metal segments, the array is well-suited to wire grid modeling using the Method of Moments (MoM).  In this study, we employ the NEC-4.1 implementation of MoM \cite{NEC4}.  

\begin{figure}
\begin{center}
%\vspace{3in}
\psfig{file=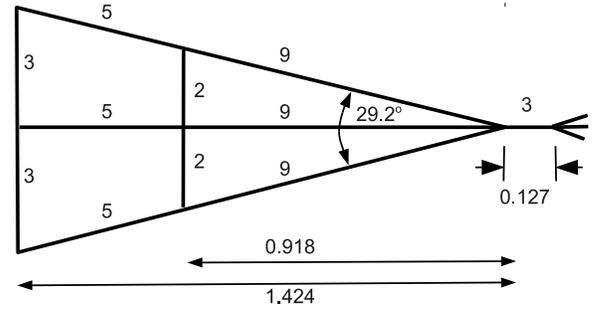,width=3.1in}
\end{center}
\caption{\label{fDipoleModel}
Geometry of the wire grid used to model the dipole. Dimensions are in meters.  The number of segments used in the MoM model are indicated next to each wire.  The radius of all wires is 1.2~cm, modeling aluminum tubing having square cross-section with 3/4-in sides.  The dipole arms are bent $45^{\circ}$ downward from the junction with the center (feed) wire.    
}
\end{figure}

The dimensions and parameters used to model the dipole are illustrated in Figure~\ref{fDipoleModel}.  The wire grid models representing each of the two antennas in a stand are vertically separated by three times the wire radius to prevent the feeds from intersecting.  The mean height of the highest points on each dipole (also the segment containing the feed) is 1.5~m above ground.  It is known from both simulations and experiment that neither the center mast nor the structure supporting the dipole arms (see Figure~\ref{fStation}) have a significant effect on the relevant properties of the dipoles, and therefore no attempt is made to model them.  It should be noted that the segmentation shown in Figure~\ref{fDipoleModel} is designed to be valid for the highest frequency of interest, so that the same model can be used for all frequencies of interest.  In order to confirm that the results were not sensitive to the selected segmentation, several alternative schemes with small changes in the number of segments per wire were also considered.  The results do not change significantly with these changes in segmentation. 

The ground screen is modeled using a 3~m $\times$ 3~m wire grid with spacing 10~cm $\times$ 10~cm and wire radius of 1~mm, which is very close to the actual dimensions.  The modeled ground screen is located 1~cm above ground to account for the significant but irregular gap that exists because of ground roughness.\footnote{Experiments with this model show that the results are not sensitive to the separation between ground screen and ground.}  The ground itself is modeled as an infinite homogeneous half-space with relative permittivity of 3 and conductivity of 100~$\mu$S, which is appropriate for ``very dry ground'' \cite{ITU_P527} which predominates in New Mexico, where LWA-1 is located.  (It should be noted that we suspect that the ground permittivity at the LWA-1 site is significantly higher; this is addressed below.)
%\footnote{Although the actual ground parameters are not known, we suspect the ground permittivity is usually higher than is assumed here, and that this results in somewhat better performance at the lowest frequencies.}  
%\footnote{It should be noted that this choice is essentially a worst case scenario in terms of sensitivity; results computed assuming}.  
Each dipole is connected to an active balun which presents a balanced input impedance of $R_L = 100~\Omega$.  For additional information on this design, the reader is referred to \cite{PIEEE_LWA} and the references therein.             

It will be useful later in this paper to know the performance of a single stand, neglecting the rest of the array.  We begin with the array manifold for the stand, which is determined as follows.  The stand is illuminated with a $\theta$-polarized 1~V/m plane wave incident from some direction $\psi$, and the resulting current $I_L$ across the series resistance modeling each active balun is determined using MoM.  Each element of the $N=2$ array response vector ${\bf a}_{\theta}(\psi)$ is then simply $I_L R_L$ for the associated antenna.\footnote{N=2 because we analyze both dipoles in a stand simultaneously.}  The process is repeated for a $\phi$-polarized plane wave and iterated over $\psi$. 

The external noise covariance matrix ${\bf P}_z$ is computed using a model proposed in \cite{E05} which assumes that Galactic noise dominates over thermal noise from the ground and other natural or anthropogenic sources of noise.  Specifically, $T_e(\psi)$ is assumed to be uniform over the sky ($\theta<\pi/2$), and zero for $\theta>\pi/2$.  In practice, $T_e(\psi)$ varies considerably both as a function of $\psi$ and as a function of time of day, due to the rotation of the Earth.  However, the above assumption provides a reasonable standard condition for comparing Galactic noise-dominated antenna systems, as explained in \cite{E05} and demonstrated in \cite{ESP07} and \cite{KL09}.  Using this model, $T_e(\psi)$ toward the sky is found to be 50,444~K, 9751~K, and 1777~K at 20~MHz, 38~MHz, and 74~MHz, respectively. The actual contributions to the system temperature are less due to the mismatch between the antenna self-impedance and $R_L$, but this is automatically taken into account as a consequence of our definition of the array manifold, which includes the loss due to impedance mismatch as well as ground loss. Under these assumptions, ${\bf P}_z$ is computed using Equation~\ref{ePzInt}.

The internal noise covariance matrix ${\bf P}_u$ is computed assuming that the internal noise associated with any given antenna is not significantly correlated with the internal noise associated with any other antenna, so that ${\bf P}_u$ becomes a diagonal matrix whose non-zero elements are:
\begin{equation}
{\bf P}_u^{\left[n\right]} = k T_{p,n} R_L 
\end{equation}
where $T_{p,n}$ is the input-referred internal noise temperature associated with the $n^{\mbox{th}}$ antenna.  We will further assume that all the electronics are identical such that $T_{p,n}=T_p$, where $T_p$ is assumed to be $250$~K, the nominal value of the cascade noise temperature of all electronics attached to a dipole, referred to the dipole terminals.    

The ratio $Tr\left\{{\bf P}_z\right\}/Tr\left\{{\bf P}_u\right\}$ (where ``$Tr$'' denotes the trace operation; i.e., the sum of the diagonal elements) is the degree to which Galactic noise dominates over internal noise in the combined output, and is found to be $-2.6$~dB, $+11.1$~dB, and $+4.1$~dB at 20, 38, and 74~MHz, respectively.  The 38~MHz and 74~MHz results are consistent with field measurements (see Figure~6 of \cite{PIEEE_LWA}), however the same measurements suggest 20~MHz should also be Galactic noise-dominated.  The apparent reason for the discrepancy is that the 20~MHz result is relatively sensitive to ground permittivity, both because the loss associated with Earth ground increases with decreasing frequency, and also because the ground screen becomes tiny (only $0.2\lambda \times 0.2\lambda$) at 20~MHz.  Larger assumed permittivity in our calculations results in Galactic noise-dominated performance at 20~MHz, even if the loss tangent is also increased.  The effect of the change of ground parameters on the 38~MHz and 74~MHz results is very small in comparison.   
%Interestingly, the SEFD results presented below are not dramatically affected by choice of ground parameters, even at 20~MHz.  
We shall continue to use the original ground parameters in this paper as they can be considered to be safely conservative.   

\begin{figure}
\begin{center}
%\vspace{3in}
\psfig{file=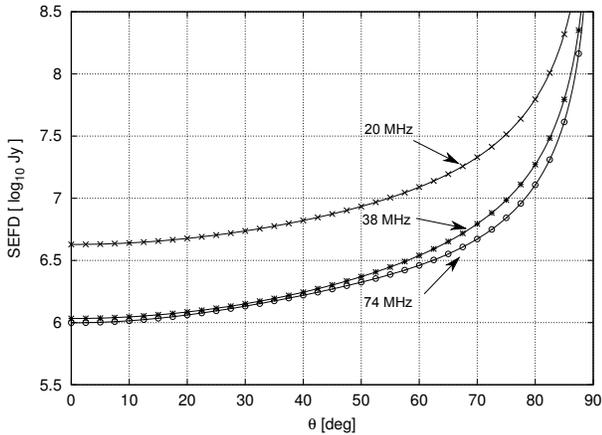,width=3.1in}
\end{center}
\caption{\label{fSingleStandSEFD}
SEFD estimates for a single LWA-1 stand (alone; the rest of the array is not present) in the $\phi=0$ half-plane. Note that lower SEFD is better (more sensitive).
}
\end{figure}

Using the array manifold and the noise covariance matrices calculated as described above, the resulting SEFD for a single stand (and neglecting the rest of the array) can be computed from Equation~\ref{eSEFD}. The result for the $\phi=0$ plane is shown in Figure~\ref{fSingleStandSEFD}.  Note Figure~\ref{fSingleStandSEFD} is also essentially a pattern measurement; as such the expected ``$\cos{\theta}$''-type behavior is evident; in particular, the response is seen to go to zero at the horizon, as expected.   Note that the performance at 38~MHz and 74~MHz is similar despite the large difference in frequency; this is because both the Galactic noise and the effective aperture of the antennas decrease with frequency at approximately the same rate \cite{PIEEE_LWA}.  The calculated 20~MHz performance is somewhat worse for the reasons described in the previous paragraph.     
%
%\begin{figure}
%\begin{center}
%%\vspace{3in}
%\psfig{file=figs/a5.eps,width=3.1in}
%\end{center}
%\caption{\label{fa5}
%Pattern for Single HiFi stand, E/H plane co-pol.
%Also shown: Result for no ground screen, and infinite/PEC. 
%% a5.m
%}
%\end{figure}

The stand performance can also be described in the traditional way, in terms of gain, through the effective aperture $A_e$.  Let the power delivered to the load ($R_L$) be $P_L$.  Note $P_L = S(\psi)A_e(\psi)$ (assuming a co-polarized incident field), and also $P_L = \left|I_L\right|^2 R_L$. Since $S(\psi) = \left|E^i(\psi)\right|^2/{\eta}$, where $E^i(\psi)$ is the co-polarized incident electric field, we have that the effective aperture for any given antenna attached to a load $R_L$ is
\begin{equation}
A_e(\psi) = \eta \frac{\left|I_L\right|^2}{\left|E^i(\psi)\right|^2} R_L~\mbox{.}
\label{eAe}
\end{equation}
Assuming that $I_L$ is computed using the MoM model described above, this definition {\it includes} impedance mismatch as well as loss due to the conductivity of the ground.\footnote{These factors can be computed independently and removed, if desired; see \cite{E05} and \cite{ESP07}.}  Using Equation~\ref{eAe}, the zenith value of $A_e$ is estimated to be 0.25~m$^2$, 8.72~m$^2$, and 2.48~m$^2$ for 20~MHz, 38~MHz, and 74~MHz, respectively for each dipole in the single-stand system described in this section.  It should be noted, however, that these values cannot be used directly to calculate a ``$A_e/T_{sys}$'' type sensitivity metric, since $T_{sys}$ in this case would be $T_e$, reduced by the impedance mismatch, plus $T_p$; and the mismatch efficiency is not available as part of this analysis.  This underscores the usefulness of SEFD as a sensitivity metric for this class of systems, in contrast to $A_e$ (or antenna gain) or $A_e/T_{sys}$.

%--------------------------------------------------------
\subsection{\label{ssArray}Computation of the LWA-1 Array Manifold}

\begin{figure}
\begin{center}
%\vspace{3in}
\psfig{file=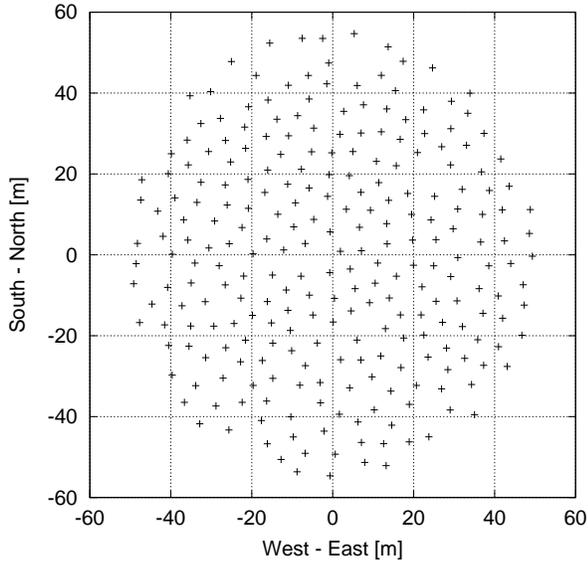,width=3.1in}
\end{center}
\caption{\label{fAG}
Arrangement of stands in the LWA-1 array.  The minimum distance between any two masts is 5~m ($0.33\lambda$, $0.63\lambda$, and $1.23\lambda$ at 20~MHz, 38~MHz, and 74~MHz, respectively).  All dipoles are aligned North-South and East-West; $\phi=0$ is East.  For additional information about the array geometry, see \cite{PIEEE_LWA}.
}
\end{figure}

The arrangement of stands in the LWA-1 array is as shown in Figure~\ref{fAG}.  We now consider the problem of modeling this array so as to obtain the array manifold.  In principle this is simply a matter of adding 255 identical stands to the model described in the previous section, and repeating the MoM analysis.  In practice, however, this leads to an intractably large model with prohibitively large computational burden.  Whereas the single stand model (including the ground screen) uses 2074 segments, the complete array so modeled would require 530,944 segments, which is well beyond the capability of commonly-available computing hardware.  A more reasonable target is a model with about 11,000 segments, which fits in 4~GB of RAM and takes 1-2 hours to run on a recent-vintage workstation-class computer.  

In this study, the number of segments used to model the array at 38~MHz and 74~MHz is reduced by performing the MoM analysis for one stand at a time, using the following procedure: (1) The present stand of interest is modeled as described in the previous section; (2) The dipoles for the remaining 255 stands are modeled using a simpler ``surrogate'' dipole, described below; and (3) The ground screens for the 19 stands closest to the stand of interest are modeled using a surrogate (sparser) wire grid, also described below, and ground screens are not included for the remaining 237 stands, under the assumption that they do not have a significant effect.   This model requires slightly fewer than 11,000 segments, and is run 256 times (once for each stand) to complete the analysis of the array at one frequency.  Analysis at one frequency requires approximately 1 month of continuous computation using a cluster of 4 computers.  The approach used for 20~MHz is the same in all respects, except a coarser grid is used for the surrogate ground screens, which allows the number of surrogate ground screens to be increased to 108.

The surrogate dipole model replaces each triangular wire grid dipole ``arm'' with a single thick wire of length 1.7235~m with radius 6~cm, which is divided into 3 segments.  This results in segment lengths of $0.038\lambda$, $0.073\lambda$, and $0.142\lambda$ at 20~MHz, 38~MHz, and 74~MHz, respectively.  This model yields nearly the same impedance vs. frequency around resonance as the original bowtie dipole.  The surrogate ground screen model increases the grid spacing to 75~cm for 20~MHz, and 30~cm for 38~MHz and 74~MHz.  These grid spacings correspond to $0.05\lambda$, $0.038\lambda$, and $0.074\lambda$ at 20~MHz, 38~MHz, and 74~MHz, respectively.  The wire radius is increased to 5~mm to compensate for the increased grid spacing while keeping the wire cross-section well clear of the Earth ground.  The required number of surrogate ground screens was determined using an experiment in which the computed characteristics of the stand of interest were observed as the number of surrogate ground screens used in surrounding stands was increased, starting with the closest stand and working outward.  It was found that ground screens within  about $1.5\lambda$ were often important, whereas ground screens for stands further away had negligible effect.  To be conservative, 19 surrogate ground screens were used for the 38~MHz and 74~MHz results, whereas 108 surrogate ground screens were used for the 20~MHz results; in each case this yields a MoM model with slightly fewer than the ``maximum manageable'' number of segments (11,000) identified above.

To further validate the array model and computation, results were computed for a subset of the dipoles in ``scaled up'' versions of the array which were identical in all respects except that the inter-stand spacings were increased.  It was confirmed that the results for any given dipole converge to the single-stand results (shown in the previous section) with sufficiently large inter-stand spacing.     

\begin{figure}
\begin{center}
%\vspace{3in}
\psfig{file=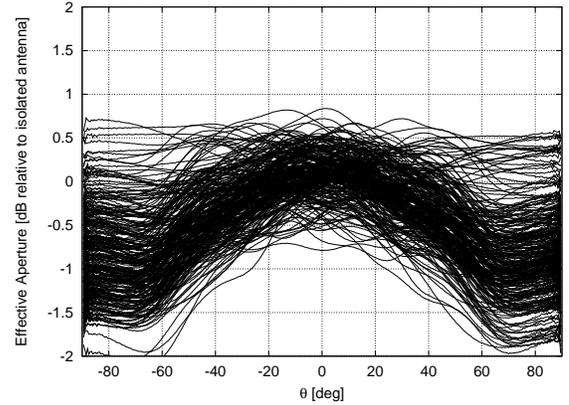,width=3.1in}
\end{center}
\caption{\label{fPats20}
$A_e(\psi)$ for all 256 North-South aligned antennas in the $\phi=0$ plane at 20~MHz, relative to the same value for a single stand in isolation.  Note this is essentially a measurement of the effect of mutual coupling on the antenna gain. 
%Note these are essentially E-plane co-polarized patterns.
% 
}
\end{figure}

\begin{figure}
\begin{center}
%\vspace{3in}
\psfig{file=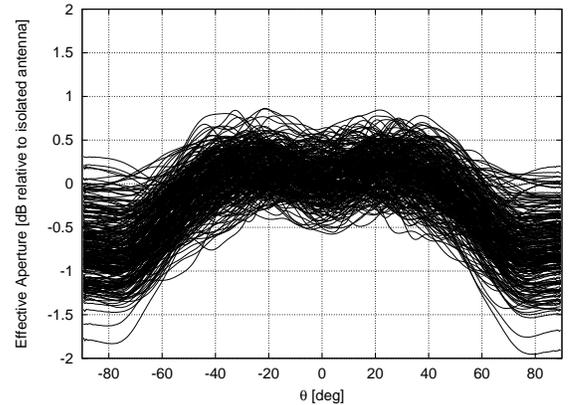,width=3.1in}
\end{center}
\caption{\label{fPats38}
Same as Figure~\ref{fPats20}, except for 38~MHz.
}
\end{figure}

\begin{figure}
\begin{center}
%\vspace{3in}
\psfig{file=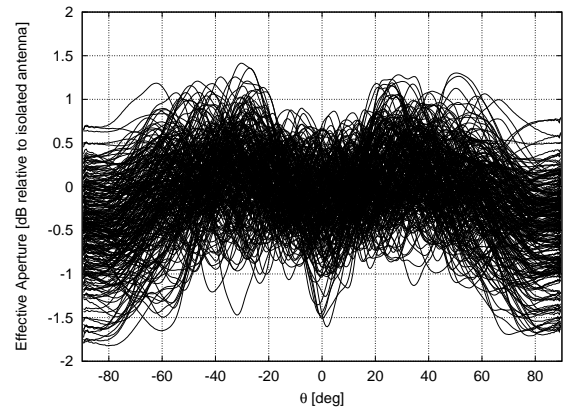,width=3.1in}
\end{center}
\caption{\label{fPats74}
Same as Figure~\ref{fPats20} and \ref{fPats38}, except for 74~MHz.
}
\end{figure}

MoM analysis reveals that the behavior of stands in the array is considerably different from stands in isolation.  This is demonstrated in Figures~\ref{fPats20}--\ref{fPats74}, which show the patterns of all 256 North-South aligned antennas in the $\phi=0$ plane at frequencies of 20~MHz, 38~MHz, and 74~MHz, respectively.
%and Figure~\ref{fa6}, which shows the feed point currents for the same antennas in response to co-polarized illumination from zenith at several frequencies.  
It is clear that the combination of non-uniform spacings and mutual coupling leads to disorderly embedded patterns.  At 20~MHz and 38~MHz, the the pattern tends to increase slightly toward the zenith, and decrease slightly more toward the horizon. At 74~MHz this trend is not as pronounced, but the pattern tends to be greater for $20^{\circ} \le \theta \le 60^{\circ}$.
%However, this is of little consequence under nominal Galactic noise-limited conditions, since both the signal of interest and the sky noise  
%

%========================================================
\section{\label{sAP}SEFD Performance of the LWA1 Array}
%========================================================  

\begin{figure}
\begin{center}
%\vspace{3in}
\psfig{file=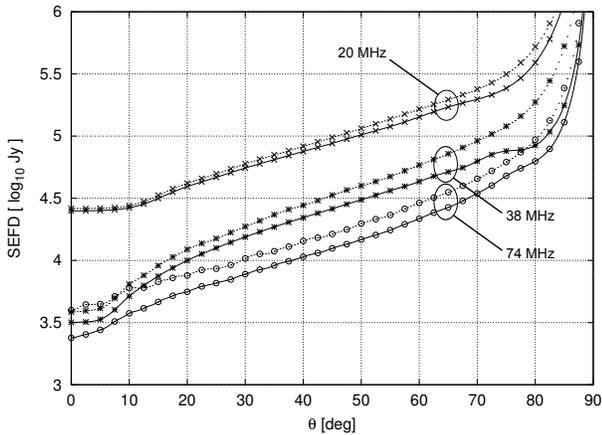,width=3.1in}
\end{center}
\caption{\label{fAP1}
Calculated SEFD of LWA-1 for beam pointing in the $\phi=0$ plane.  For each frequency, the upper (dotted) curve is the result for simple beamforming, and the lower (solid) curve is the result for optimal beamforming.
% a41.m
}
\end{figure}

Using the currents obtained as described in Section~\ref{ssArray}, it is possible to calculate the SEFD as defined in Equation~\ref{eSEFD}.  Figure~\ref{fAP1} shows results in the $\phi=0$ plane.  For each frequency, results are shown for two beamforming schemes: (1) ``simple'' beamforming, in which the coefficients ({\bf b}) are determined by geometrical delays (and no other considerations); and (2) ``optimal'' (maximum SNR) beamforming, in which ${\bf b}$ is chosen to be the eigenvector associated with the largest eigenvalue of ${\bf R_n}^{-1}{\bf R}_s$ (as discussed in Section~\ref{sTheory}).  As expected, optimal beamforming consistently outperforms simple beamforming, with the typical improvement being in the range 1--2~dB.
%; for example, SEFD for the zenith-pointing beam is improved by about x~dB, 0.8~dB, and x~dB at 20~MHz, 38~MHz, and 74~MHz, respectively.   
%

\begin{figure}
\begin{center}
%\vspace{3in}
\psfig{file=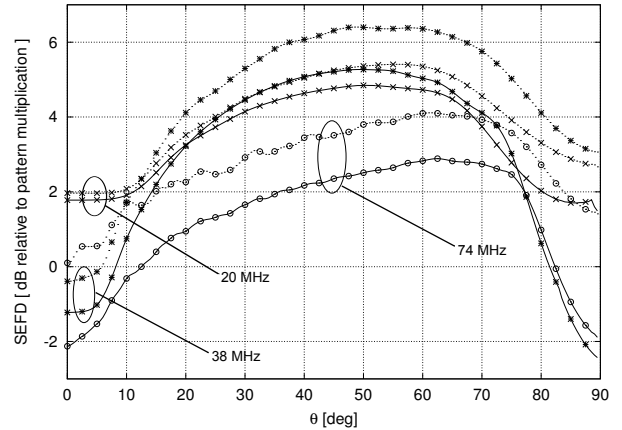,width=3.1in}
\end{center}
\caption{\label{fAP1delta}
The result from Figure~\ref{fAP1} divided by the result expected from pattern multiplication (i.e., the SEFD from Figure~\ref{fSingleStandSEFD}, divided by 256).  
}
\end{figure}

\begin{figure}
\begin{center}
%\vspace{3in}
\psfig{file=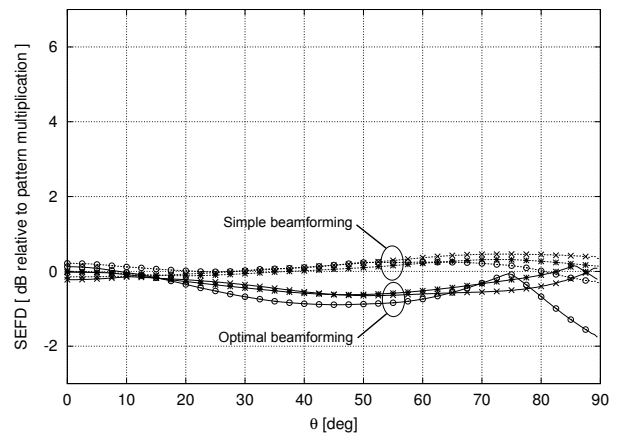,width=3.1in}
\end{center}
\caption{\label{fAP2}
Same as Figure~\ref{fAP1delta}, except computed with external (Galactic) noise correlation ``turned off'' (see text).  For $\theta<75^{\circ}$, the results for 20~MHz, 38~MHz, and 74~MHz are very close.
% a41.m
}
\end{figure}

If the principle of pattern multiplication applies, then we would expect the Figure~\ref{fAP1} result to be identical to the Figure~\ref{fSingleStandSEFD} result (for the single stand in isolation), scaled by the number of stands (256).  However, this is not the case, as is shown in Figure~\ref{fAP1delta}. The SEFD is greater (i.e., worse) than the result predicted by pattern multiplication by about 1--6~dB (varying with frequency and $\theta$) for $\theta$ greater than about $20^{\circ}$, and is different (not consistently better or worse) for $\theta$ less than about $20^{\circ}$.  Two possible culprits are mutual coupling and Galactic noise correlation.  From Section~\ref{ssArray} it is clear that mutual coupling is significant.  However, the primary culprit is Galactic noise correlation, as demonstrated in Figure~\ref{fAP2}.  This figure shows a recalculation of the Figure~\ref{fAP1delta} result with ${\bf P}_z^{\left[n,m\right]}$ set to zero for $n \neq m$; i.e., forcing the correlation of the external noise received by different antennas to be zero.  This yields a result which is relatively close to that predicted by pattern multiplication; thus correlation of external noise between antennas is primarily responsible for the reduced sensitivity.     

Given the large effect mutual coupling is seen to have on individual antenna patterns, it is interesting that the results of simple beamforming should be so close to the pattern multiplication results.  Also interesting is the finding that optimum beamforming still provides a benefit of about 1~dB at all frequencies for $20^{\circ} < \theta < 75^{\circ}$, even with external noise correlation ``turned off''.  Mutual coupling is, in this sense, beneficial; although optimum beamforming coefficients are required to realize the benefit. 

\begin{figure}
\begin{center}
%\vspace{3in}
\psfig{file=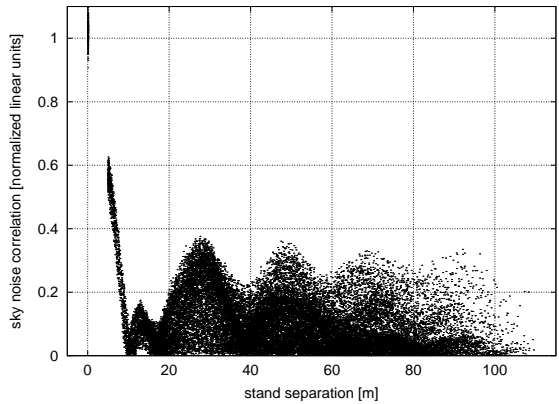,width=3.1in}
\end{center}
\caption{\label{fa31_20}
The magnitude of the sky noise correlation ${\bf P}_z^{\left[n,m\right]}$ at 20~MHz as a function of separation between stands $n$ and $m$, shown as a scatter plot where each point corresponds to one pair of stands.  The results have been normalized so that the maximum value (i.e., for $n=m$) assuming pattern multiplication is unity (1). Note the minimum spacing between the masts of any two stands is 5~m; thus no points exist for spacings greater than zero and less than 5~m.  
% a6.m
}
\end{figure}

\begin{figure}
\begin{center}
%\vspace{3in}
\psfig{file=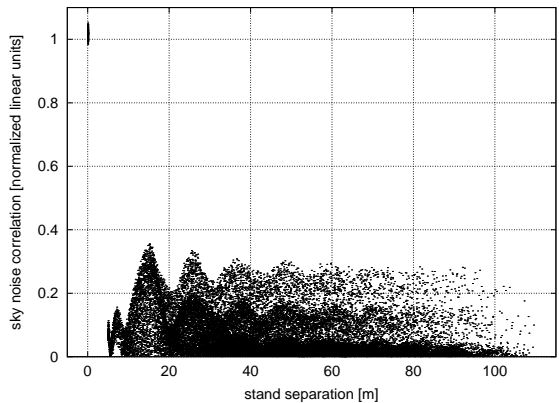,width=3.1in}
\end{center}
\caption{\label{fa31_38}
Same as Figure~\ref{fa31_20}, except at 38~MHz.
% a6.m
}
\end{figure}

\begin{figure}
\begin{center}
%\vspace{3in}
\psfig{file=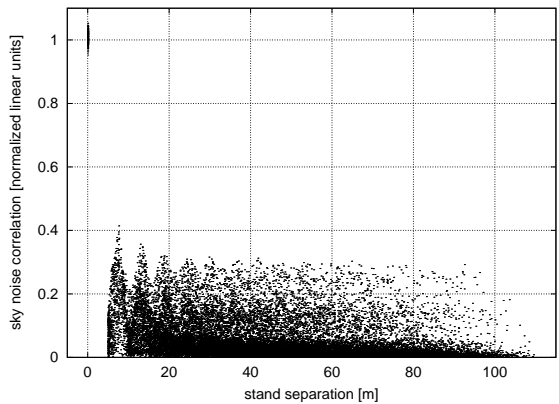,width=3.1in}
\end{center}
\caption{\label{fa31_74}
Same as Figure~\ref{fa31_20} and \ref{fa31_38}, except at 74~MHz.
% a6.m
}
\end{figure}

Further insight can be gained from Figures~\ref{fa31_20}--\ref{fa31_74}, which show that Galactic noise correlation is quite large for closely-spaced stands, and in many cases is large even for antennas on opposite sides of the array.  Thus, it is not surprising that sensitivity tends to be degraded relative to a similar calculation in which external noise correlation is assumed to be zero.  It is interesting to note that the correlation exhibits a Bessel function-like trend as a function of separation in wavelengths.  However, it should be emphasized that this result assumes uniform sky brightness, and (as pointed out earlier) the actual situation is somewhat different.  Non-uniform sky brightness will introduce structure in the external noise covariance matrix (${\bf P}_z$) that is likely to cause corresponding $\psi$-dependent variations in SEFD.

%These findings are remarkable for two reasons.  First, it implies that the mutual coupling, despite the fact that it is strong enough to significantly distort individual antenna patterns, has very little effect on beamforming performance -- even if it is ignored for the purposes of computing the beamforming coefficients. 
%Second, it implies that sky noise correlation significantly degrades the sensitivity of the array, and that even using the best possible beamforming coefficients, the array sensitivity remains significantly less than would be predicted by pattern multiplication.  

%========================================================
\section{\label{sConc}Conclusions}
%========================================================

This paper has considered the sensitivity of large arrays of low-gain antenna elements at low frequencies for which Galactic noise can be an important or dominant part of the system temperature.
%, which includes a number of new and planned radio telescopes as well as riometers and HF- and VHF-band phased array radars.  
General expressions were developed for SNR (Equation~\ref{eSNR}) and SEFD (Equation~\ref{eSEFD}) for beamforming in terms of the array manifold and internal and external covariance matrices.  Some results are shown using LWA-1 at 20~MHz, 38~MHz, and 74~MHz as an application example.  It is shown that for beams pointing more than $10^{\circ}$--$20^{\circ}$ away from the zenith, the combination of mutual coupling and correlation of Galactic noise between antennas results in sensitivity which is significantly worse than predicted by pattern multiplication beginning with single antennas in isolation.  Closer to the zenith, the result is frequency-dependent, and can be better or worse than the result predicted by pattern multiplication.  It is also shown that improvement of 1--2~dB is possible by using beamforming coefficients specifically designed to maximize SNR, as opposed to coefficients derived solely from geometrical phase and which therefore neglect external noise correlation as well as mutual coupling.      

The ultimate intended use of LWA-1 is not solely as a stand-alone instrument, but rather as one of 53 identical ``stations'' distributed over the State of New Mexico which are combined to form images using aperture synthesis techniques \cite{PIEEE_LWA}.  Because the minimum separation between stations will be on the order of kilometers, the effects of mutual coupling and spatial correlation of Galactic noise will be negligible in the process of combining station beams into an image.  Thus, the SEFD for imaging will be better by a factor of $\sqrt{N_S(N_S-1)}$ than the SEFD for the station beam, where $N_S$ is the number of stations.  Adopting a value of 3200~Jy for the typical zenith-pointing SEFD from Figure~\ref{fAP1}, the SEFD for imaging near the zenith with $N_S=53$ is expected to be about 61~Jy.  The resulting near-zenith image sensitivity, assuming 1~h integration, 8~MHz bandwidth, and SNR$=5$, is about 2~mJy.  This is consistent with the result derived in \cite{PIEEE_LWA}, 
%which is based on pattern multiplication with a empirically-derived constant margin used to account for mutual coupling.  
which neglected Galactic noise correlation.
However results for imaging at larger zenith angles will not be consistent with \cite{PIEEE_LWA}, for the reasons discussed above.  Better estimates for pointing directions in the $\phi=0$ plane can be obtained starting with Figure~\ref{fAP1}.      

Finally, it should be noted that the theory and techniques described in Section~\ref{sTheory} are generally applicable; even to arrays employing regular spacings, with or without mutual coupling, and dominated or not by external noise.  Other findings in this paper may be specifically relevant for arrays used in other radio science applications, including HF/VHF direction finding arrays, radar arrays for measuring the atmosphere or ionosphere, and riometers.

%====================================================================
%====================================================================
% use section* for acknowledgement

\section*{Acknowledgments}
% optional entry into table of contents (if used)
%\addcontentsline{toc}{section}{Acknowledgment}

The LWA is supported by the Office of Naval Research through a contract
with the University of New Mexico.  In addition to the efforts of the many members of the many institutions involved in the LWA project, the author acknowledges specific contributions to the design of the LWA-1, as related in this paper, from 
A.\ Cohen,  
B.\ Hicks,  
N.\ Paravastu,  
P.\ Ray, and
H.\ Schmidt 
of the U.S. Naval Research Laboratory;
S.\ Burns 
of Burns Industries; and 
J.\ Craig and
S.\ Tremblay 
of the University of New Mexico.

\begin{thebibliography}{1}

\bibitem{PIEEE_LWA} %2%
S.W.\ Ellingson, T.E.\ Clarke, A.\ Cohen, N.E.\ Kassim, Y.\ Pihlstr\"{o}m, L.\ J Rickard, and G.B.\ Taylor, ``The Long Wavelength Array,'' {\it Proc. IEEE}, Vol.~97, No.~8, pp. 1421--30, Aug 2009. 

\bibitem{PIEEE_LOFAR} %1%
M.\ de Vos, A.W.\ Gunst, and R.\ Nijboer, ``The LOFAR Telescope: System Architecture and Signal Processing,'' {\it Proc. IEEE}, Vol.~97, No.~8, pp. 1431--7, Aug 2009. 

\bibitem{PIEEE_MWA} %3%
C.J.\ Lonsdale {\it et al.}, ``The Murchison Widefield Array: Design Overview,'' {\it Proc. IEEE}, Vol.~97, No.~8, pp. 1497--1506, Aug 2009. 

\bibitem{PIEEE_SKA} %4%
P.E.\ Dewdney, P.J.\ Hall, R.T.\ Schilizzi, \& T.J.L.W.\ Lazio, ``The Square Kilometre Array,'' {\it Proc. IEEE}, Vol.~97, No.~8, pp. 1482--96, Aug 2009.                    

\bibitem{E05} %5%
S.W. Ellingson, ``Antennas for the Next Generation of Low-Frequency Radio Telescopes,'' {\it IEEE Trans. Antennas \& Prop.}, Vol.~53, No.~8, pp.~2480--9, Aug 2005.

\bibitem{VB88} %6%
B.D.\ Van Veen \& K.M.\ Buckley, ``Beamforming: A Versatile Approach to Spatial Filtering,'' {\it IEEE ASSP Magazine}, pp. 4--24, Apr 1988.
%ASSPMag8804_VanVeenBuckley.pdf

\bibitem{Lee93} %7%
J.J.\ Lee, ``$G/T$ and Noise Figure of Active Array Antennas,'' {\it IEEE Trans. Antennas \& Propagation}, Vol.~41, No.~2, pp. 241--4, Feb 1993.
% 00214619.pdf
% Early attempt to characterize sensitivity of arrays in terms of G/T. -- not so useful for low frequency radio astronomy...?

\bibitem{Kraft00} %8%
U.R. Kraft, ``Gain and $G/T$ of Multielement Receive Antennas with Active Beamforming Networks,'' {\it IEEE Trans. Antennas \& Propagation}, Vol.~48, No.~12, pp. 1818--29, Dec 2000.
% 00901270.pdf
% Subsequent attempt to characterize sensitivity of arrays in terms of G/T. -- not so useful for low frequency radio astronomy...?

\bibitem{VBA05} %9%
W.A.\ Van Cappellen, J.D.\ Bregman, and M.J.\ Arts, ``Effective Sensitivity of a Non-Uniform Phased Array of Short Dipoles,'' {\it Experimental Astronomy}, Vol.~17, pp. 101--109, 2004. 
% vanCappellen_sparseArrays05.pdf
% Neglects sky noise correlation

\bibitem{LWA142} %10%
S.\ Ellingson, ``Sky Noise-Induced Spatial Correlation,'' Memo 142, Oct 14, 2008, Long Wavelength Array Memo Series [Online]. Available: http://www.phys.unm.edu/$\sim$lwa/memos.

\bibitem{CPD04} %11%
C.\ Craeye, B.\ Parvais, \& X. Dardenne, ``MoM Simulation of Signal-to-Noise Patterns in Infinite and Finite Receiving Antenna Arrays,'' {\it IEEE Trans. Antennas \& Propagation}, Vol.~52, No.~12, pp.~3245--56, Dec 2004.
%TAP0412_Craeye_MOMSimOfSNPatterns.pdf
%Did MoM, S/N analysis; but uniform arrays, internal-noise limited arrays. Main issue is finite vs. infinite array (not pattern multiplication.)

\bibitem{C05} %12%
C.\ Craeye, ``Including Spatial Correlation of Thermal Noise in the Noise Model of High-Sensitivity Arrays,'' {\it IEEE Trans. Antennas \& Propagation}, Vol.~53, No.~11, pp.~3845--48, Nov 2005.
%TAP0511_Craeye_IncludingSpatialCorrelation.pdf
% spatial correlation is due to lossy antennas.

\bibitem{IMW08} %13%
M.V.\ Ivashina, R.\ Maaskant, \& B.\ Woestenburg, ``Equivalent System Representation to Model the Beam Sensitivity of Receiving Antenna Arrays,'' {\it IEEE Ant. \& Wireless Prop.\ Let.}, Vol.~7, pp.~733--7, 2008.
% AWPL_ArrayModel.pdf

\bibitem{Gans06} %14%
M.J.\ Gans, ``Channel Capacity Between Antenna Arrays -- Part I: Sky Noise Dominates,'' {\it IEEE Trans. Communications}, Vol.~54, No.~9, pp.~1586--92, Sep 2006.
%01703816_Gans_ChCapSkyNoise.pdf

\bibitem{AST07} %15%
A.\ Kisliansky, R.\ Shavit, \& J.\ Tabrikian, ``Direction of Arrival Estimation in the Presence of Noise Coupling in Antenna Arrays,'' {\it IEEE Trans. Antennas \& Prop.}, Vol.~55, No.~7, pp.~1940--7, July 2007.
%TAP0707_DOAEst_in_NoiseCoupling.pdf

\bibitem{LS67}
Y.T.\ Lo \& R.J. Simcoe, ``An Experiment on Antenna Arrays with Randomly Spaced Elements,'' {\it IEEE Trans. Antennas \& Prop.}, Vol.~AP--15, No.~2, Mar 1967, pp.~231--5.
% TAP67_LoSimcoe_RandomArray.pdf

\bibitem{AL72}
V.D.\ Agrawal \& Y.T.\ Lo, ``Mutual Coupling in Phased Arrays of Randomly Spaced Antennas,'' {\it IEEE Trans. Antennas \& Prop.}, Vol.~AP--20, No.~3, May 1972, pp.~288--95.
% TAP72_AgrawalLo.pdf
                                    
\bibitem{MM80} %16%
R.A. Monzingo and T.W. Miller, {\it Introduction to Adaptive Arrays}, Wiley, 1980.  (Reprinted by SciTech Publishing, 2004.)

\bibitem{NEC4} %17%
G.J.\ Burke, {\it Numerical Electromagnetics Code -- NEC-4, Method of Moments}, Lawrence Livermore National Laboratory Technical Report UCRL-MA-109338, Parts I--III, Jan 1992. 

\bibitem{ITU_P527} %18%
International Telecommunications Union, ``Electrical Characteristics of the Surface of the Earth,'' Recommendation P.527--3, 1992. 

\bibitem{ESP07} %19%
S.W. Ellingson, J.H. Simonetti, and C.D. Patterson, ``Design and Evaluation of an Active Antenna for a 29-47 MHz Radio Telescope Array,'' {\it IEEE Trans. Antennas \& Prop.}, Vol.~55, No.~3, Mar 2007, pp. 826--31.

\bibitem{KL09} %20%
A.\ Kerkhoff \& H.\ Ling, ``Design of Broadband Antenna Elements for a Low-Frequency Radio Telescope using Pareto Genetic Algorithm Optimization,'' {\it Radio Science}, Vol.~44, RS6006, doi:10.1029/2008RS004131, 2009.

%\bibitem{LWAweb}
%Long Wavelength Array project website [on-line] {http://lwa.unm.edu}.

\end{thebibliography}
\end{document}